\documentstyle[amssymb,preprint,aps]{revtex}

\begin{document}
\title{Fifth force from fifth dimension: a comparison between two different
approaches}
\author{F. Dahia$^{a}$, E. M. Monte$^{b}$ and C. Romero$^{b}$}
\address{$^{a}$Departamento de F\'{i}sica, Universidade Federal de Campina Grande,\\
58109-970, Campina Grande, Pb, Brazil\\
$^{b}$Departamento de F\'{i}sica, Universidade Federal da Para\'{i}ba, C.\\
Postal 5008, 58051-970 Jo\~{a}o Pessoa, Pb, Brazil\\
e-mail: cromero@fisica.ufpb.br}
\maketitle

\begin{abstract}
We investigate the dynamics of particles moving in a spacetime augmented by
one extra dimension in the context of the induced matter theory of gravity.
We examine the appearance of a fifth force as an effect caused by the extra
dimension and discuss two different approaches to the fifth force formalism.
We then give two simple examples of application of both approaches by
considering the case of a Ricci-flat warped-product manifold and a
generalized Randall-Sundrum space.
\end{abstract}

\section{ \ \ \ Introduction}

\begin{quote}
The idea that our ordinary spacetime might be embedded in a
higher-dimensional manifold seems to pervade almost all the recent
development of Cosmology and Particle Physics theory. From the old
Kaluza-Klein\cite{kaluza,klein} model to modern theories of supergravity and
superstrings \cite{collins,appel} the assumption that extra dimensions do
exist, though not observed yet, appears to be closely connected with the
belief that all forces of nature are ultimately different aspects of a
single entity. Along with the search for unification there is another
motivation for constructing higher-dimensional theories which goes back to
Einstein and consists in regarding the physical world as a manifestation of
pure geometry\cite{Wheeler}. Of these two schemes the latter includes the
so-called induced-matter theory (IMT) or non-compactified Kaluza-Klein
theory of gravity, an approach which regards macroscopic matter as being
geometrically ``induced'' by a mechanism that locally embeds our
four-dimensional (4D) spacetime in a Ricci-flat five-dimensional manifold \ 
\cite{Wesson,book,Overduin,Ponce,Liu}. As we know, the original version of
Kaluza-Klein theory assumes as a postulate that the fifth dimension is
compact. However, in the case of the induced-matter theory (IMT) this
requirement has been dropped. Moreover, it is asserted that only one extra
dimension should be sufficient to explain all the phenomenological
properties of matter. More specifically, IMT proposes that the classical
energy-momentum tensor, which enters the right-hand side of the Einstein
equations could be, in principle, generated by pure geometrical means. In
other words, geometrical curvature would induce matter in four dimensions
and to observers in the physical 4D spacetime the extra dimension would
appear as the matter source for gravity. One interesting point is that the
matter ``generated'' by this process is of a very general kind, i.e. {\it %
any }energy-momentum tensor can be produced by choosing the appropriate
embedding, a result which is mathematically supported by a powerful theorem
of differential geometry due to Campbell and Magaard \cite
{campbell,magaard,romero,seahra2}. On the other hand, the question of how
the dynamics of test particles is influenced by the fifth dimension has been
examined recently \cite{Wesson2,Seahra,Ponce2,Mashhoon}. It has been argued
that departures from conventional 4D dynamics could be interpreted as due to
the existence of a fifth force, such deviations being in principle amenable
to measurement. It seems, however, that the mathematical prescription to
calculate this force is not unique and, in fact, depends on basic
assumptions of the five-dimensional underlying \ theory. It is our purpose
here to examine this point in a more critical way and give alternative
definitions of the concept of fifth force. As we shall see, these concepts
depend strictly on whether the geometry of the four-dimensional observed
spacetime is defined by an embedding mechanism or by a foliation of the
five-dimensional (5D) manifold.
\end{quote}

\section{\protect\bigskip The axioms of the induced-matter theory}

\ Let us briefly spell out what seems to be the basic axioms or postulates
implicitly assumed in the induced-matter theory or non-compactified
Kaluza-Klein gravity, as it sometimes is referred to in its original
formulation.

i) Our ordinary space-time $M^{4}$ may be represented as a four-dimensional
hypersurface $\Sigma ^{4}$ locally and isometrically embedded in a
Ricci-flat differentiable manifold $M^{5}$.

Now it is well known that the five-dimensional line element$\
ds^{2}=g_{ab}dx^{a}dx^{b}$ of $M^{5}$ can always be put, at least locally,
in the form (see, for example,\cite{magaard}) 
\begin{equation}
ds^{2}=g_{\alpha \beta }dx^{\alpha }dx^{\beta }+\epsilon \phi ^{2}d\psi ^{2}
\label{foliation}
\end{equation}
where $x^{a}=(x^{\alpha },\psi )$, $g_{\alpha \beta }=g_{\alpha \beta
}(x^{a})$, $\phi =\phi (x^{a})$ and $\epsilon =\pm 1$, depending on whether
the fifth dimension is chosen timelike or spacelike. (Throughout Latin
indices take value in the range (0,1,...,4) and Greek indices run from
(0,1,2,3)). We now assume that $M^{4}$ is identified with the hypersurface $%
\Sigma ^{4}$ defined by the equation $\psi =$ \ \ $\psi _{0}=$constant.
Then, we have an induced metric on $\Sigma ^{4}$ given by 
\begin{equation}
^{(4)}g_{\alpha \beta }(x^{\mu })=g_{\alpha \beta }(x^{\mu },\psi _{0})
\label{inducedmetric}
\end{equation}
On the other hand, in terms of the 5D Christoffel symbols the 5D Ricci
tensor is given by 
\begin{equation}
R_{ab}=(\Gamma _{ab}^{c})_{,c}-(\Gamma _{ac}^{c})_{,b}+\Gamma
_{ab}^{c}\Gamma _{cd}^{d}-\Gamma _{ad}^{c}\Gamma _{bc}^{d}
\label{riccitensor}
\end{equation}
Putting $a\rightarrow \alpha ,b\rightarrow \beta $ in the above equation
will give us the 4D components of the 5D Ricci tensor. It is not difficult
to verify that we can rewrite (\ref{riccitensor}) as 
\begin{equation}
R_{\alpha \beta }=^{(4)}R_{\alpha \beta }+\Gamma _{\alpha \beta
,4}^{4}-\Gamma _{\alpha 4,\beta }^{4}+\Gamma _{\alpha \beta }^{\lambda
}\Gamma _{\lambda 4}^{4}+\Gamma _{\alpha \beta }^{4}\Gamma _{4D}^{D}-\Gamma
_{\alpha \lambda }^{4}\Gamma _{\beta 4}^{\lambda }-\Gamma _{\alpha
4}^{D}\Gamma _{\beta D}^{4}
\end{equation}
where $^{(4)}R_{\alpha \beta }$ may be looked upon as the 4D Ricci tensor
calculated with the metric $^{(4)}g_{\alpha \beta }(x^{\mu })$ provided that
in the above equation we set $\psi =$ \ \ $\psi _{0}$. Finally, we can show
that the 5D vacuum Einstein equations $G_{ab}=0$ can be written, separetely,
in the following way: 
\begin{equation}
^{(4)}G_{\alpha \beta }=\kappa ^{(4)}T_{\alpha \beta }
\end{equation}
where $^{(4)}T_{\alpha \beta }$ is now interpreted as the energy-momentum
tensor of ordinary 4D matter, and is given explicitly by

\bigskip 
\begin{equation}
^{(4)}T_{\alpha \beta }=\frac{\phi _{\alpha ;\beta }}{\phi }-\frac{%
\varepsilon }{2\phi ^{2}}\left[ \frac{\phi _{,4}g_{\alpha \beta ,4}}{\phi }%
-g_{\alpha \beta ,44}+g^{\lambda \mu }g_{\alpha \lambda ,4}g_{\beta \mu ,4}-%
\frac{g^{\mu \nu }g_{\mu \nu ,4}g_{\alpha \beta ,4}}{2}+\frac{g_{\alpha
\beta }}{4}\left\{ g_{,4}^{\mu \nu }g_{\mu \nu ,4}+\left( g^{\mu \nu }g_{\mu
\nu ,4}\right) ^{2}\right\} \right]  \label{energy-momentum tensor}
\end{equation}

\begin{equation}
\varepsilon \phi \square \phi =-\frac{g_{,4}^{\lambda \beta }g_{\lambda
\beta ,4}}{4}-\frac{g^{\lambda \beta }g_{\lambda \beta ,44}}{2}+\frac{\phi
,_{4}g^{\lambda \beta }g_{\lambda \beta ,4}}{2\phi }  \label{scalarfield}
\end{equation}
which may be viewed as an equation for a scalar field $\phi ;$ and

\begin{equation}
P_{\alpha ;\beta }^{\beta }=0
\end{equation}
an equation that has the appearance of a conservation law, where $P_{a\beta
} $ is defined by 
\begin{equation}
P_{\alpha \beta }=\frac{1}{2\sqrt{g_{44}}}\left( g_{\alpha \beta
,4}-g_{\alpha \beta }g^{\mu \nu }g_{\mu \nu ,4}\right)  \label{palfabeta}
\end{equation}

\bigskip\ We now state a second postulate:

ii)\ The energy-momentum tensor which describes the matter content of the
four-dimensional Universe will be given by the equation (\ref
{energy-momentum tensor}).

\qquad \qquad \qquad

For completion of the theory we need a third postulate concerning the motion
of free-falling test particles and light rays. This point will be dealt with
in the next section.

\bigskip

\section{ The fifth force in the induced-matter theory}

\bigskip

\bigskip The induced-matter theory has often been regarded in the literature
as an embedding theory, i.e. a theory which \ assumes as a first principle
that our ordinary spacetime corresponds to some hypersurface embedded in
some higher-dimensional manifold ( in this case, a 5D Ricci-flat space). In
the same sense, the recently proposed brane-world theory \cite
{randall-sundrum}\ may also be called an embedding theory since the brane
which models our observable Universe is viewed as a four-dimensional
hypersurface embedded in a five-dimensional anti-de Sitter manifold (the
so-called bulk). ( The relationship between the induced-matter and
brane-world theories is discussed in ref.\cite{Ponce3} ).

It turns out, however, that when it comes to the dynamics of test particles
in IMT it has been implicitly assumed that the paths of these particles
correspond to curves in the five-dimensional manifold $M^{5}$, not
necessarily confined to the hypersurface $\Sigma ^{4}$. In this respect let
us recall that in the brane-world model of the Universe matter and radiation
are confined to the brane, although in high energy regime particles, as well
as gravitons, can in principle leave the brane. In this case the effects of
the motion in the fifth dimension are expected to appear as a force
affecting the motion of particles in four dimensions.

\section{Two approaches to the fifth force}

Let us now consider two distinct approaches to the fifth force which will be
referred to as the{\it \ foliating }and {\it \ embedding approaches}. They
are defined as follows:

i) The foliating approach makes use of a congruence of a given vector field $%
V$ defined in $M^{5}$ and implicitly assumes that the equations governing
the 4D observed physical laws are in a way ''projections'' of 5D equations
onto a foliation of hypersurfaces $\left\{ \Sigma \right\} $ orthogonal to $%
V.$ In this approach the fifth force is determined by inducing the metric of 
$M^{5}$ on the leaves.

ii) In the embedding approach it is also assumed that the fundamental 5D
manifold $M^{5}$ can be foliated by a set of \ hypersurfaces $\left\{ \Sigma
\right\} $ orthogonal to a vector field $V.$ However, here the geometry of
the 4D space-time is not supposed to be determined by the entire foliation,
but by a particular leaf $\Sigma ^{4}$ selected from the set $\left\{ \Sigma
\right\} $, on which a metric tensor is induced by the embedding manifold $%
M^{5}.$ In \ this approach the fifth force is determined in terms of
geometrical quantities which are defined exclusively in $\Sigma ^{4}.$

We then add a third postulate concerning the motion of particles and light:

{\it The paths corresponding to the motion of free-falling test particles
and light rays are geodesic lines in the 5D fundamental Ricci-flat space }$%
M^{5}$.

\section{The 5D dynamics in the foliating approach}

\ \bigskip The distinction made in the previous section between two possible
approaches to the induced-matter theory is crucial when one attempts at
defining which has been called a ''fifth force'' acting on test particles.
Traditionally this has been done by examining how the geodesic equations in
5D splits up when they are ''projected'' ortogonally onto the leaves of the
foliation defined by the vector field\ $\frac{\partial }{\partial \psi }$ 
\cite{Wesson2}. As we shall see, the equation of the projected geodesics
contains a term which can be viewed as a ``force''. It turns out that this
force depends in general on the 4D components of the 5D Christoffel symbols $%
\Gamma _{bc}^{a}(x^{a},\psi )$, which are {\it not }to be identified with
the Christoffel symbols $^{(4)}\Gamma _{bc}^{a}(x^{a},\psi _{0})$ of the 4D
hypersurface $\Sigma ^{4}$ (these latter are calculated with the induced
metric $\ ^{(4)}g_{\alpha \beta }(x^{\mu })=g_{\alpha \beta }(x^{\mu },\psi
_{0})$ defined on $\Sigma ^{4}$). Thus, in the the traditional approach (in
accordance with the third postulate), the motion of particles and light rays
is assumed to be governed by the five-dimensional geodesic equation:

\bigskip

\begin{equation}
\frac{d^{2}x^{a}}{dS^{2}}+\Gamma _{bc}^{a}\frac{dx^{b}}{dS}\frac{dx^{c}}{dS}%
=0  \label{geodesics}
\end{equation}
where $S$ is an affine parameter.

Let us now show that when the equation of motion of a particle is augmented
by one dimension, extra terms may appear which can have an interesting
interpretation as an anomalous acceleration or a force (per unit mass). (One
would plausibly argue here that to detect deviations of 4D general
relativity dynamics could provide an indirect way of testing the space-time
dimensionality). In order to simplify the dynamics we set $\phi =1$ in the
equation (\ref{foliation}), which amounts to disregard the effects of a
possible scalar field in the usual interpretation of Kaluza-Klein theory 
\cite{Liu}.\bigskip\ With respect to the congruence defined by $V$ let us
define the projector

\begin{equation}
h_{ab}(x,\psi )=g_{ab}(x,\psi )-V_{a}V_{b}  \label{projection}
\end{equation}

This projector allow us to split the five-dimensional metric $g_{ab}(x,\psi
) $ into a part parallel to $\frac{\partial }{\partial \psi }$ and a part
orthogonal to the leaves of the foliation defined by $\psi =const$. We thus
define the 4D metric on each leaf by 
\begin{equation}
^{(4)}\hat{g}_{\mu \nu }(x,\psi )\equiv h_{\mu }^{c}h_{\nu
}^{d}g_{cd}(x,\psi )=g_{\mu \nu }(x,\psi )  \label{projected metric}
\end{equation}
\bigskip \qquad\ In this approach all the geometric quantities which are
relevant in 4D are to be constructed from $^{(4)}\hat{g}_{\mu \nu }(x,\psi
), $ and not from the induced metric $g_{\alpha \beta }(x^{\mu },\psi _{0})$%
. Clearly, the mathematical formalism developed in Section 2 for the case of
an embedding theory, which leads to the equations (4-9), is exactly the same
for the case of a foliating theory. Our aim now is to look at the
five-dimensional geodesic equation (\ref{geodesics}) as being constituted of
an equation of motion in four dimensions plus an equation for the fifth
coordinate $\psi $. We shall first consider timelike curves in 5D and also
assume that $\epsilon =-1$, although these restrictions are by no means
essential, i.e. null curves may also be considered . Thus, given a 5D
timelike curve $x^{a}=x^{a}(\lambda )$ we define the 5D proper time function 
$S=S(\lambda )$ by 
\begin{equation}
S(\lambda )=\int_{\lambda _{0}}^{\lambda }\left[ g_{ab}\frac{dx^{a}}{d\tau }%
\frac{dx^{b}}{d\tau }\right] ^{1/2}d\tau =\int_{\lambda _{0}}^{\lambda }%
\left[ g_{\mu \nu }\frac{dx^{\mu }}{d\tau }\frac{dx^{\nu }}{d\tau }-\left( 
\frac{d\psi }{d\tau }\right) ^{2}\right] ^{1/2}d\tau  \label{5Dpropertime}
\end{equation}
where $\lambda $ is a real parameter and $\lambda _{0}$ an arbitrary
constant. Since we are assuming a positive integrand in the above equation $%
S $ has inverse, $\lambda =\lambda (S),$ and we have the equation

$\label{derivada}$%
\begin{equation}
\frac{d\lambda }{dS}=\left[ g_{\mu \nu }\frac{dx^{\mu }}{d\lambda }\frac{%
dx^{\nu }}{d\lambda }-\left( \frac{d\psi }{d\lambda }\right) ^{2}\right]
^{-1/2}  \label{derivadaS}
\end{equation}
\qquad \qquad\ \ \qquad \qquad \qquad \qquad In the same way, for the same
curve $x^{a}=x^{a}(\lambda )$ we define the 4D proper time function by 
\begin{equation}
s(\lambda )=\int_{\lambda _{0}}^{\lambda }\left[ g_{\mu \nu }\frac{dx^{\mu }%
}{d\tau }\frac{dx^{\nu }}{d\tau }\right] ^{1/2}d\tau  \label{4Dpropertime}
\end{equation}
from which we have 
\begin{equation}
\frac{d\lambda }{ds}=\left[ g_{\mu \nu }\frac{dx^{\mu }}{d\lambda }\frac{%
dx^{\nu }}{d\lambda }\right] ^{-1/2}  \label{derivadas}
\end{equation}

\bigskip Here we make two important assumptions: i) although the world-line
of a particle is a curve in 5D what we directly observe is its ``4D part '' $%
x^{\mu }=x^{\mu }(\lambda )$; ii) accordingly, $s(\lambda )$ gives the
proper time ellapsed along that curve as measured by a 4D clock. Note that $%
s(\lambda )$ is calculated with the 4D components of 5D metric $%
g_{ab}(x,\psi )$, i.e. $g_{\mu \nu }(x,\psi ),$ not being restricted to any
particular hypersurface of the foliation.

By taking $\lambda =s$ we deduce from (\ref{derivadaS}) and (\ref{derivadas}%
) the following equation 
\begin{equation}
\frac{ds}{dS}=\left[ 1-\left( \frac{d\psi }{ds}\right) ^{2}\right] ^{-1/2}
\label{reparametrization}
\end{equation}

Using (\ref{reparametrization}) to reparametrise (\ref{geodesics}) we easily
obtain the equations below 
\begin{equation}
\frac{d^{2}x^{\mu }}{ds^{2}}+\Gamma _{\alpha \beta }^{\mu }\frac{dx^{\alpha }%
}{ds}\frac{dx^{\beta }}{ds}=-\left( \frac{d^{2}s}{dS^{2}}\right) \left( 
\frac{ds}{dS}\right) ^{-2}\frac{dx^{\mu }}{ds}-2\Gamma _{4\nu }^{\mu }\frac{%
dx^{\nu }}{ds}\frac{d\psi }{ds}  \label{fifthforce}
\end{equation}
\begin{equation}
\frac{d^{2}\psi }{ds^{2}}+\Gamma _{\alpha \beta }^{4}\frac{dx^{\alpha }}{ds}%
\frac{dx^{\beta }}{ds}=-\left( \frac{d^{2}s}{dS^{2}}\right) \left( \frac{ds}{%
dS}\right) ^{-2}\frac{d\psi }{ds}  \label{equationforpsi}
\end{equation}

\bigskip In the foliating approach the right-hand side of equation (\ref
{fifthforce}), which depends on the fifth coordinate $\psi $ is interpreted
as a fifth force (per unit mass). In terms of the 5D metric components this
quantity may be expressed as

\begin{equation}
f^{\mu }=-\frac{dx^{\nu }}{ds}\frac{d\psi }{ds}\left( \left[ 1-\left( \frac{%
d\psi }{ds}\right) ^{2}\right] ^{-1}\delta _{\nu }^{\mu }\frac{d^{2}\psi }{%
ds^{2}}+g^{\mu \alpha }\frac{\partial g_{\alpha \nu }}{\partial \psi }\right)
\label{fifthforcefoliating}
\end{equation}
It follows immediately that if the 5D curve lies entirely in a hypersurface $%
\psi =const$, then no fifth force arises in this formulation.

\section{The 5D dynamics in the embedding approach}

In the embedding approach our 4D spacetime is regarded as the
four-dimensional hypersurface $\Sigma ^{4}$ defined by $\psi =\psi
_{0}=const $. Therefore all physical quantities which in principle can be
measured in 4D should be expressed in terms of the coordinates $x^{\mu }$
(which cover the hypersurface $\Sigma ^{4})$. According to this view a curve
\ $x^{a}=x^{a}(\lambda )$ lying on $M^{5}$ is accessible to four-dimensional
observers through its projection onto $M^{4}$, $x^{\mu }=x^{\mu }(\lambda )$%
. On the other hand, as we have pointed out in Section II, the geometry of $%
M^{4}$ is to be identified to the geometry of the hypersurface $\Sigma ^{4}$%
, defined by $\psi =\psi _{0},$ with a metric tensor given by 
\begin{equation}
^{(4)}g_{\alpha \beta }(x^{\mu })=g_{\alpha \beta }(x^{\mu },\psi _{0})
\end{equation}
Thus the Christoffel symbols of $M^{4}$ are to be calculated with the metric
above.

Therefore one has to consider the 4D projection of the 5D geodesic equation (%
\ref{geodesics}) , which will be given by

\begin{equation}
\frac{d^{2}x^{\mu }}{dS^{2}}+\Gamma _{bc}^{\mu }\frac{dx^{b}}{dS}\frac{dx^{c}%
}{dS}=0  \label{4dgeodesics}
\end{equation}
where now the 4D proper time $\ $%
\begin{equation}
\sigma (\lambda )=\int_{\lambda _{0}}^{\lambda }\left[ ^{(4)}g_{\mu \nu }%
\frac{dx^{\mu }}{d\tau }\frac{dx^{\nu }}{d\tau }\right] ^{1/2}d\tau
\label{temposigma}
\end{equation}
must be calculated with the metric $^{(4)}g_{\mu \upsilon }(x^{\alpha
})=g_{\mu \upsilon }(x^{\alpha },\psi _{0})$. Note that in this approach
both the metric tensor $^{(4)}g_{\alpha \beta }$ and the proper time $\sigma 
$ are quantities which are perfectly measurable in $\Sigma ^{4}$ and do not
depend on the fifth coordinate $\psi $. If we also define $\Delta \Gamma
_{\alpha \beta }^{\mu }=\Gamma _{\alpha \beta }^{\mu }(x^{\nu },\psi
)-^{(4)}\Gamma _{\alpha \beta }^{\mu }(x^{\nu },\psi _{0})$, where $%
^{(4)}\Gamma _{\alpha \beta }^{\mu }$ $(x^{\nu },\psi _{0})$ denotes the
Christoffel symbols of $M^{4}$ calculated with $^{(4)}g_{\mu \upsilon
}(x^{\alpha })$, then the equation (\ref{4dgeodesics}) can be put in the form

\begin{equation}
\frac{d^{2}x^{\mu }}{d\sigma ^{2}}+^{(4)}\Gamma _{\alpha \beta }^{\mu }\frac{%
dx^{\alpha }}{d\sigma }\frac{dx^{\beta }}{d\sigma }=-\left( \frac{%
d^{2}\sigma }{dS^{2}}\right) \left( \frac{d\sigma }{dS}\right) ^{-2}\frac{%
dx^{\mu }}{d\sigma }-2\Gamma _{4\nu }^{\mu }\frac{dx^{\nu }}{d\sigma }\frac{%
d\psi }{d\sigma }-\Delta \Gamma _{\alpha \beta }^{\mu }\frac{dx^{\alpha }}{%
d\sigma }\frac{dx^{\beta }}{d\sigma }  \label{fifthforceembedding}
\end{equation}
Clearly, the left-hand side of this equation corresponds to the absolute
derivative $\frac{DV^{\mu }}{Ds}$ of the vector $V^{\mu }$ tangent to the 4D
projected curve $x^{\mu }=x^{\mu }(\sigma )$. Hence it would be natural to
consider, in the embedding approach, the fifth force (per unit mass) as
being given by

\begin{equation}
f^{\mu }=-\left( \frac{d^{2}\sigma }{dS^{2}}\right) \left( \frac{d\sigma }{dS%
}\right) ^{-2}\frac{dx^{\mu }}{d\sigma }-2\Gamma _{4\nu }^{\mu }\frac{%
dx^{\nu }}{d\sigma }\frac{d\psi }{d\sigma }-\Delta \Gamma _{\alpha \beta
}^{\mu }\frac{dx^{\alpha }}{d\sigma }\frac{dx^{\beta }}{d\sigma }
\end{equation}
or, in terms of the metric components, 
\begin{equation}
f^{\mu }=-\frac{dx^{\nu }}{d\sigma }\frac{d\psi }{d\sigma }\left( \left[
1-\left( \frac{d\psi }{d\sigma }\right) ^{2}\right] ^{-1}\delta _{\nu }^{\mu
}\frac{d^{2}\psi }{d\sigma ^{2}}+g^{\mu \alpha }(x,\psi _{0})\left( \frac{%
\partial g_{\alpha \nu }}{\partial \psi }\right) _{\psi =\psi _{0}}\right)
-\Delta \Gamma _{\alpha \beta }^{\mu }\frac{dx^{\alpha }}{d\sigma }\frac{%
dx^{\beta }}{d\sigma }
\end{equation}

\qquad Likewise, the equation governing the motion in the extra dimension
will be given by 
\begin{equation}
\frac{d^{2}\psi }{d\sigma ^{2}}+^{(4)}\Gamma _{\alpha \beta }^{4}\frac{%
dx^{\alpha }}{d\sigma }\frac{dx^{\beta }}{d\sigma }=-\left( \frac{%
d^{2}\sigma }{dS^{2}}\right) \left( \frac{d\sigma }{dS}\right) ^{-2}\frac{%
d\psi }{d\sigma }-\Delta \Gamma _{\alpha \beta }^{4}\frac{dx^{\alpha }}{%
d\sigma }\frac{dx^{\beta }}{d\sigma }
\end{equation}

\section{ Applications of the two formalisms}

In this section we wish to work out the equations obtained in the preceding
sections through two examples: the case of a Ricci-flat warped-product
manifold and a generalized Randall-Sundrum space. Our aim is to compare the
results obtained for the fifth force as prescribed by the foliating and
embedding approaches.

Let us firstly consider the five-dimensional metric 
\begin{equation}
dS^{2}=\Lambda \frac{\psi ^{2}}{3}\left( dt^{2}-e^{2\sqrt{\frac{\Lambda }{3}}%
t}\left[ dx^{2}+dy^{2}+dz^{2}\right] \right) -d\psi ^{2}  \label{deSitter}
\end{equation}

\bigskip This is a 5D Ricci-flat manifold which may be viewed as an
embedding space for the 4D de Sitter space-time with the metric 
\begin{equation}
d\sigma ^{2}=dt^{2}-e^{2\sqrt{\frac{\Lambda }{3}}t}\left[
dx^{2}+dy^{2}+dz^{2}\right]
\end{equation}
\bigskip the embedding taking place at the branes $\psi =\pm \psi _{0}=\pm 
\sqrt{\frac{3}{\Lambda }}$\cite{Ponce}\cite{romero}. The nonvanishing
Christoffel symbols calculated from (\ref{deSitter}) are $\Gamma
_{04}^{0}=1/\psi $, $\Gamma _{11}^{0}=$ $\Gamma _{22}^{0}=\Gamma _{33}^{0}=%
\sqrt{\Lambda /3}e^{2\sqrt{\frac{\Lambda }{3}}t}$, $\Gamma _{0j}^{i}=\delta
_{j}^{i}\sqrt{\Lambda /3}$, $\Gamma _{j4}^{i}=\delta _{j}^{i}/\psi $, $%
\Gamma _{00}^{4}=\Lambda \psi /3$, $\Gamma _{11}^{4}=\Gamma _{22}^{4}=\Gamma
_{33}^{4}=-\Lambda /3\psi e^{2\sqrt{\frac{\Lambda }{3}}t}$ $(i=1,2,3).$

Thus, if $x^{a}=x^{a}(\lambda )$ is the curve corresponding to the worldline
of a test particle, then the 5D proper time function $S=S(\lambda )$ will be
given by 
\begin{equation}
S(\lambda )=\int_{\lambda _{0}}^{\lambda }\left[ \Lambda \frac{\psi ^{2}}{3}%
\left( (dt/d\tau )^{2}-e^{2\sqrt{\frac{\Lambda }{3}}t}\left[ (dx/d\tau
)^{2}+(dy/d\tau )^{2}+(dz/d\tau )^{2}\right] \right) -(d\psi /d\tau )^{2}%
\right] ^{1/2}d\tau
\end{equation}

\bigskip In the foliating and embedding approaches the 4D proper time
functions will be given, respectively, by 
\begin{equation}
s(\lambda )=\int_{\lambda _{0}}^{\lambda }\left[ \Lambda \frac{\psi ^{2}}{3}%
\left( (dt/d\tau )^{2}-e^{2\sqrt{\frac{\Lambda }{3}}t}\left[ (dx/d\tau
)^{2}+(dy/d\tau )^{2}+(dz/d\tau )^{2}\right] \right) \right] ^{1/2}d\tau
\label{tempopróprioS}
\end{equation}
\begin{equation}
\sigma (\lambda )=\int_{\lambda _{0}}^{\lambda }\left[ \left( (dt/d\tau
)^{2}-e^{2\sqrt{\frac{\Lambda }{3}}t}\left[ (dx/d\tau )^{2}+(dy/d\tau
)^{2}+(dz/d\tau )^{2}\right] \right) \right] ^{1/2}d\tau
\label{tempoprópriosigma}
\end{equation}

\bigskip Our next step is to obtain the geodesic equations for the 5D warped
geometry (\ref{deSitter}). These will be given by

\begin{equation}
\frac{d^{2}t}{dS^{2}}+\frac{2}{\psi }\frac{dt}{dS}\frac{d\psi }{dS}+\left(
\Lambda /3\right) ^{1/2}e^{2\left( \Lambda /3\right) ^{1/2}t}\left[
(dx/dS)^{2}+(dy/dS)^{2}+(dz/dS)^{2}\right] =0  \label{geodésicat}
\end{equation}

\bigskip 
\begin{equation}
\frac{d^{2}x^{i}}{dS^{2}}+\frac{2}{\psi }\frac{dx^{i}}{dS}\frac{d\psi }{dS}%
+2\left( \Lambda /3\right) ^{1/2}\frac{dt}{dS}\frac{dx^{i}}{dS}=0,\text{ }%
(i=1,2,3)\text{ }  \label{geodésicax}
\end{equation}

\bigskip 
\begin{equation}
\frac{d^{2}\psi }{dS^{2}}+(\Lambda \psi /3)(dt/dS)^{2}-\psi \left( \Lambda
/3\right) e^{2\left( \Lambda /3\right) ^{1/2}t}\left[
(dx/dS)^{2}+(dy/dS)^{2}+(dz/dS)^{2}\right] =0  \label{geodésicapsi}
\end{equation}

\bigskip

\qquad\ \ A particular solution of the above system is the curve whose
parametric equations are 
\begin{equation}
t=(1/2)\sqrt{3/\Lambda }\log (aS+b)  \label{soluçãot}
\end{equation}
\begin{equation}
x^{i}=c^{i}=const  \label{soluçãox}
\end{equation}
\begin{equation}
\psi =\sqrt{aS+b}  \label{soluçãopsi}
\end{equation}
where $a,b$ and $c^{i}$ are integration constants. It is easy to verify that
the curve above is a null geodesic in 5D. As it has been pointed out by some
authors\cite{sanjeev} , null geodesics in 5D may appear as timelike curves
in 4D, which would suggest interpreting the rest mass of particles in terms
of a fifth dimension. Surely the fact that in the case of null curves the
affine parameter $S$ can no longer be interpreted as proper time in 5D does
not affect the formulation of fifth force theory in both approaches, except
that now the equation (\ref{derivadaS}) is not well defined. In this case
one must work with (\ref{4Dpropertime}) and (\ref{temposigma}), which are
perfectly well defined.

\qquad Let us now calculate the fifth force prescribed by the foliating
approach , that is, from the equation (\ref{fifthforcefoliating}),(or,
equivalently, directly from the right-hand side of (\ref{fifthforce})) and (%
\ref{soluçãot}), (\ref{soluçãox}), (\ref{soluçãopsi}). A straightforward
calculation leads to the following expression for $f^{\mu }$: 
\begin{equation}
f^{0}=-\sqrt{3/\Lambda }(aS+b)^{-1},\text{ }f^{i}=0,\text{ }i=1,2,3
\end{equation}

\bigskip \qquad On the other hand, the fifth force according to the
embedding approach may be evaluated from the equations (\ref
{fifthforceembedding}) plus (\ref{soluçãot}), (\ref{soluçãox}), (\ref
{soluçãopsi}). It is easy to verify that in this case 
\begin{equation}
f^{\mu }=0,\text{ }\mu =0,1,2,3
\end{equation}

\qquad Therefore from the equations above we see that the fifth force
clearly depends on which approach is chosen. Let us also note that one can
express the components of the fifth force in terms of $s$ or $\sigma $ (the
4D proper time in the foliation or in the embedding approach, respectively).
This is simply done with the help of the equations (\ref{4Dpropertime}) and (%
\ref{temposigma}).

\qquad As a second example let us consider a space with a generalized
Randall-Sundum metric given by 
\begin{equation}
ds^{2}=e^{F(\psi )}(dt^{2}-dx^{2}-dy^{2}-dz^{2})-d\psi ^{2}
\label{randall-sundrumg}
\end{equation}
where $F(\psi )$ is an arbitrary warp factor. Clearly the above manifold can
be foliated by a set $\left\{ \Sigma \right\} $ of hypersurfaces $\psi =const
$, each leaf corresponding to a 4D Minkowski space-time. If we extend the
framework of the induced-matter theory to include the case when the
embedding 5D space is an Einstein space\cite{fabio}, i.e. a manifold in
which $R_{ab}=\Lambda g_{ab}$ \ (Ricci-flat is a particular case of an
Einstein space when $\Lambda =0$), then clearly the formalism that defines
the fifth force can be kept unaltered in both approaches. The same is true
if we consider a more general embedding space such as the one given by (\ref
{randall-sundrumg}). Let us now show that for the latter (which includes the
Randall-Sundrum as a particular case) the foliating approach leads to a
nonzero fifth force, while if we employ the embedding approach particles
lying on a brane fell no fifth force at all.

\qquad Let us consider 5D timelike geodesics in the geometry given by (\ref
{randall-sundrumg}). The timelike geodesic equations can be readily put in
the form 
\begin{equation}
\frac{dx^{\alpha }}{dS}=e^{-F(\psi )}a^{\alpha }  \label{aalfa}
\end{equation}
\begin{equation}
\left( \frac{d\psi }{dS}\right) ^{2}=e^{-F(\psi )}\eta _{\alpha \beta
}a^{\alpha }a^{\beta }-1  \label{beta}
\end{equation}
where \ by $a^{\alpha }$ we denote four integration constants and $\eta
_{\alpha \beta }$ stands for the Minkowski metric tensor.

\qquad It is easily seen that the equations relating the 5D proper time with
the 4D proper time in both the foliating and embedding approaches can be
written, respectively, as 
\begin{equation}
\frac{dS}{ds}=\frac{e^{F(\psi )/2}}{\sqrt{a^{\alpha }a_{\alpha }}}
\label{s1}
\end{equation}
and 
\begin{equation}
\frac{dS}{d\sigma }=\frac{e^{F(\psi )}}{\sqrt{a^{\alpha }a_{\alpha }}}
\label{sigma1}
\end{equation}
where we are using the convention $a_{\alpha }=\eta _{\alpha \beta }a^{\beta
}$ and, as the integration constants are arbitrary, they can be chosen such
that $a^{2}\equiv a^{\alpha }a_{\alpha }$ $>0$.

\qquad Let us now calculate the fifth force in the foliating approach by
considering the left-hand side of the equation (\ref{fifthforce}). Since $%
\Gamma _{\alpha \beta }^{\mu }=0$ we have 
\begin{equation}
f^{\mu }=\frac{d^{2}x^{\mu }}{ds^{2}}+\Gamma _{\alpha \beta }^{\mu }\frac{%
dx^{\alpha }}{ds}\frac{dx^{\beta }}{ds}=\frac{d^{2}x^{\mu }}{ds^{2}}=-\frac{%
a^{\mu }F^{\prime }}{2a^{2}}\left[ a^{2}e^{-F(\psi )}-1\right] ^{1/2}
\label{force1}
\end{equation}
where $F^{\prime }=\frac{dF}{d\psi }$.

\qquad In a similar manner we obtain, taking the embedding approach, 
\begin{equation}
f^{\mu }=\frac{d^{2}x^{\mu }}{d\sigma ^{2}}+^{(4)}\Gamma _{\alpha \beta
}^{\mu }\frac{dx^{\alpha }}{d\sigma }\frac{dx^{\beta }}{d\sigma }=\frac{%
d^{2}x^{\mu }}{d\sigma ^{2}}=0  \label{force2}
\end{equation}
since in this case we have $\frac{dx^{\mu }}{d\sigma }=\frac{a^{\mu }}{a}%
=const$ and $^{(4)}\Gamma _{\alpha \beta }^{\mu }=0$.

\qquad Therefore we conclude that in the case of the Randall-Sundrum model a
fifth force may, in principle, arise if one follows the foliating approach.
On the other hand, if one assumes the embedding approach, no fifth force can
be detected on the brane.

\section{Final comments}

\bigskip

\qquad As far as the methodology employed in this work is concerned two
comments are in order. Firstly, it would be interesting to work out a more
general formulation of the fifth force  which takes into account the usual
terms corresponding to scalar and electromagnetic fields as in Kaluza-Klein
theories. In this respect, some significant results have been achieved
basically in the framework of the foliating approach\cite{Ponce4}. Secondly,
it should be added that the character of the extra dimension (timelike or
spacelike) does not seem to be an essential part of the formalisms discussed
in this paper. Thus, both approaches may be applied, for instance, to
investigate the fifth force in the context of five-dimensional relativity
models with two times, a recent subject of research\cite{wesson4}.

\bigskip

\section{Acknowledgements}

\bigskip

C. Romero would like to thank CNPq (Brazil) for financial support.


\bigskip

\bigskip

\bigskip

\bigskip

\begin{references}
\bibitem{kaluza}  T. Kaluza, Sitz. Preuss. Akad. Wiss. 33 (1921) 966

\bibitem{klein}  O. Klein, Z. Phys. 37 (1926) 895

\bibitem{collins}  P. Collins, A. Martin and E. Squires, Particle Physics
and Cosmology, Wiley, New York, 1989

\bibitem{appel}  T. Appelquist, A. Chodos and P. Freund, Modern Kaluza-Klein
Theories, Addison-Wesley, Menlo Park, 1989

\bibitem{Wheeler}  J. A. Wheeler, Einstein%
\'{}%
s Vision, Springer, Berlin, 1968

\bibitem{Wesson}  P. S. Wesson , J. Ponce de Leon, J. Math. Phys. 33 (1992)
3883

\bibitem{book}  P. S. Wesson, Space-Time-Matter, World Scientific,
Singapore, 1999

\bibitem{Overduin}  J. M. Overduin, P. S. Wesson, Phys. Rep. 283 (1997) 303

\bibitem{Ponce}  J. Ponce de Leon, Gen. Rel. Grav. 20 (1988) 539

\bibitem{Liu}  P. S. Wesson, B. Mashhoon, H. Liu, W. N. Sajko, Phys. Lett. B
456 (1999) 34

\bibitem{campbell}  J. E. Campbell, A Course of Differential Geometry,
Clarendon, 1926

\bibitem{magaard}  L. Magaard, Zur Einbettung Riemannscher Raume in
Einstein-raume und konform-euclidische Raume, PhD Thesis, Kiel, 1963

\bibitem{romero}  C. Romero, R. Tavakol, R. Zalaletdinov, Gen. Rel. Grav. 28
(1996), 365

\bibitem{seahra2}  S. S. Seahra, P. S. Wesson, gr-qc/0302015 (2003)

\bibitem{Wesson2}  B. Mashhoon, P.S. Wesson, H.Liu, \ Gen. Rel. Grav. 30
(1998) 555

\bibitem{Seahra}  S. S. Seahra, Phys. Rev. D 65 (2002) 124004

\bibitem{Ponce2}  J. Ponce de Leon, Phys. Lett. B 523 (2001) 311

\bibitem{Mashhoon}  P. S. Wesson, B. Mashhoon, H. Liu, Mod. Phys. Lett. A 12
(1997) 2309

\bibitem{randall-sundrum}  L. Randall and R. Sundrum, Phys. Rev. Lett. 83
(1999) 3370

\bibitem{Ponce3}  J. Ponce de Leon, Mod. Phys. Lett A 16 (2001) 2291

\bibitem{sanjeev}  S. S. Seahra, P. S. Wesson, Gen. Rel. Grav. 33 (2001) 1731

\bibitem{fabio}  F. Dahia, C. Romero, J. Math. Phys. 43 (2002) 5804

\bibitem{Ponce4}  P. S. Wesson and J. Ponce de Leon, Astron. Astrophys.294
(1995) 1

\bibitem{wesson4}  P.S. Wesson, Phys. Lett. B 538 (2002) 159
\end{references}
\end{document}